\documentclass[12pt]{extarticle}
\pdfoutput=1 
\usepackage{slashed}
\usepackage{xcolor}
\usepackage{latexsym}
\usepackage{amssymb}
\usepackage{amsmath}
\usepackage{graphicx}
\usepackage{arydshln}
\usepackage{adjustbox}
\usepackage{easybmat}
\usepackage{bbm}
\usepackage{cite}
\usepackage{slashed}
\usepackage{bm}
\usepackage{adjustbox}
\usepackage{booktabs}
\usepackage{multirow} 
\usepackage{rotating}
\usepackage{mathtools}
\usepackage{lscape}
\usepackage{booktabs}
\usepackage{hyperref}
\usepackage{bm}
\usepackage{calrsfs}
\newcommand{\ii}{\ensuremath{\mathrm{i}}}

\DeclareMathOperator*{\sumint}{
\mathchoice
 {\ooalign{$\displaystyle\sum$\cr\hidewidth$\displaystyle\int$\hidewidth\cr}}
 {\ooalign{\raisebox{.14\height}{\scalebox{.7}{$\textstyle\sum$}}\cr\hidewidth$\textstyle\int$\hidewidth\cr}}
 {\ooalign{\raisebox{.2\height}{\scalebox{.6}{$\scriptstyle\sum$}}\cr$\scriptstyle\int$\cr}}
 {\ooalign{\raisebox{.2\height}{\scalebox{.6}{$\scriptstyle\sum$}}\cr$\scriptstyle\int$\cr}}
}

\begin{document}
\begin{flushright}
DESY-25-045
\end{flushright}
\thispagestyle{empty}
\begin{flushright}
\end{flushright}
\vspace{0.8cm}

\begin{center}
{\Large\sc The High-Temperature Limit of the SM(EFT)}
\vspace{0.8cm}

\textbf{
Mikael Chala$^a$ and Guilherme Guedes$^b$
}\\
\vspace{1.cm}
{\em $^a$ {Departamento de F\'isica Te\'orica y del Cosmos, Universidad de Granada, Campus de Fuentenueva, E--18071 Granada, Spain}}\\[0.3cm]
{\em $^b$ {Deutsches Elektronen-Synchrotron DESY, Notkestr. 85, 22607 Hamburg, Germany}}
\vspace{0.5cm}
\end{center}
\begin{abstract}
We derive the one-loop effective 3-dimensional Lagrangian that describes the high-temperature limit of the electroweak theory, to order $\mathcal{O}(g^6)$ in coupling constants $g$, including corrections due to Matsubara modes of both fermionic and bosonic degrees of freedom. We clarify certain aspects of the gauge-independence of physical parameters.  We also extend the calculation to the Standard Model effective field theory, paving the way, in particular, for a precise study of the electroweak phase transition within this framework.
\end{abstract}

\newpage


\section{Introduction}
The modern study of equilibrium phenomena within quantum field theory at high temperature relies on the so called \textit{dimensional reduction} (DR) formalism~\cite{Ginsparg:1980ef,Appelquist:1981vg}. This approach exploits the hierarchy of scales between the temperature, $T$, and the masses, $m$, of light fields to split the dynamics of a given system into simpler single-scale problems. In practice, effective field theory (EFT) methods are employed; the high-mass Matsubara modes~\cite{Matsubara:1955ws} arising upon compactification of the time dimension of the full 4-dimensional theory over a radius $R\sim 1/T$ are integrated out and their effects are captured by the Wilson Coefficients (WC) of an Euclidean 3-dimensional EFT involving only the zero modes of bosonic fields~\cite{Braaten:1995cm,Kajantie:1995dw}.

Advantages of this approach include, among others, (i) a well-defined local effective action even 
when evaluated at momentum-dependent field configurations~\cite{Strumia:1998nf,Berges:1996ib,Croon:2020cgk}; (ii) a better convergence of perturbation theory near phase transition (PT) temperatures, with limited dependence on the renormalization scale~\cite{Croon:2020cgk,Gould:2021oba,Gould:2023ovu}; (iii) large-logarithms resummation, $\sim \log{T/m}$, using renormalization group techniques~\cite{Farakos:1994kx}; (iv) the possibility of simulating the non-perturbative dynamics on the lattice~\cite{Farakos:1994xh,Kajantie:1995kf,Laine:1995np,Laine:1997dy,Gurtler:1997hr,Rummukainen:1998as,Laine:1998jb,Moore:2000jw,Arnold:2001ir,Sun:2002cc,DOnofrio:2015gop,Gould:2022ran}.
DR techniques have been widely applied to the study of hot QCD~\cite{Braaten:1994na,Braaten:1995cm,Braaten:1995jr,Kajantie:1997tt,Laine:2019uua,Laine:2018lgj,Ghiglieri:2021bom}; more recently, they are being utilized with increasing interest, precision and sophistication, in the computation of first-order PT parameters~\cite{Brauner:2016fla,
Andersen:2017ika,Niemi:2018asa,Gorda:2018hvi,Kainulainen:2019kyp,Gould:2019qek,Niemi:2020hto,Gould:2021ccf,Gould:2021dzl,Schicho:2021gca,Lofgren:2021ogg,Niemi:2021qvp,Niemi:2022bjg,Ekstedt:2022ceo,Gould:2022ran,Ekstedt:2022zro,Biondini:2022ggt,Schicho:2022wty,Gould:2023jbz,Kierkla:2023von,Aarts:2023vsf,Niemi:2024axp,Chala:2024xll,Qin:2024idc,Gould:2024jjt,Niemi:2024vzw,Kierkla:2025qyz}. This endeavor has been boosted significantly in view of current~\cite{LIGOScientific:2014pky,NANOGrav:2020bcs} and future~\cite{Harry:2006fi,Kawamura:2006up,Ruan:2018tsw,Caprini:2019egz} observatories of gravitational waves (GW) produced during this sort of PTs, smoking guns of physics beyond the Standard Model (SM).

Extensions of the SM studied within DR include the singlet~\cite{Schicho:2021gca,Brauner:2016fla,Niemi:2021qvp,Niemi:2024axp,Niemi:2024vzw}, doublet~\cite{Andersen:2017ika,Gorda:2018hvi,Kainulainen:2019kyp} and triplet~\cite{Niemi:2018asa} extensions and, more recently, the SM EFT~\cite{Croon:2020cgk,Camargo-Molina:2021zgz,Camargo-Molina:2024sde}; see also Ref.~\cite{Ekstedt:2022bff} for a tool aimed at computing the 3-dimensional EFT of arbitrary models of weakly-coupled particles. 
Most of these works concentrate on the effects of the leading interactions in the 3-dimensional EFT, often neglecting higher-dimensional operators.
However, recent studies have showed evidence that higher-order corrections~\cite{Chapman:1994vk,Moore:1995jv,Laine:2018lgj,Chakrabortty:2024wto} can be very important for accurately describing
the signatures of very strong PTs and the resulting GWs~\cite{Chala:2024xll,Bernardo:2025vkz}. (Moreover, the breaking of certain symmetries, e.g. charge-parity (CP), only manifests itself in high-order EFT terms~\cite{Kajantie:1997ky}.)

Hence, in this article we compute the one-loop next-to-leading-order terms in the 3-dimensional EFT of the electroweak (EW) sector of the SM, which includes operators of up to dimension 6 in 4-dimensional units.  We incorporate the effects of bosonic Matsubara modes, thus extending previous calculations where only the top quark was considered~\cite{Moore:1995jv}. We also clarify certain issues related to the gauge-dependency of dimension-6 interactions, pointed out in Ref.~\cite{Croon:2020cgk}. Furthermore, we go beyond the SM by including dimension-6 SMEFT interactions in the full 4-dimensional theory, hence complementing recent works towards understanding the phase structure of this theory~\cite{Camargo-Molina:2021zgz,Camargo-Molina:2024sde}; see also Refs.~\cite{Zhang:1992fs,
Bodeker:2004ws,Grojean:2004xa,Ham:2004zs,Delaunay:2007wb,
Grinstein:2008qi,Damgaard:2015con,deVries:2017ncy,Cai:2017tmh,
Chala:2018ari,Ellis:2018mja,Ellis:2019oqb,Ellis:2019tjf,Phong:2020ybr,Postma:2020toi,Hashino:2022ghd,
Kanemura:2022txx,Alonso:2023jsi,Oikonomou:2024jms
}. Our results might be also relevant for precise computations of other thermal parameters, e.g. the SM pressure at high temperature~\cite{Gynther:2005dj}.

The article is organized as follows. We 
introduce our conventions in section~\ref{sec:conventions}, where we also describe precisely the full set of independent Green's functions necessary for the matching computations of our work. We provide our main findings in section~\ref{sec:results}, which includes the matching corrections to all 3-dimensional parameters, resulting from equating off-shell correlators computed within both the 4-dimensional and 3-dimensional theories. We dedicate section~\ref{sec:crosschecks} to discuss different cross-checks of our results, including comparison to previous calculations in the literature as well as more formal aspects such as gauge-independence of physical parameters. We close in section~\ref{sec:outlook}, where we also comment on an immediate application of our work and on future directions. 

\section{Conventions}
\label{sec:conventions}
Our main goal is deriving the 3-dimensional effective theory describing the high-temperature limit of the EW sector of the SM as well as of its extension with dimension-6 operators, also known as SMEFT~\cite{Buchmuller:1985jz,Grzadkowski:2010es,Brivio:2017vri,Isidori:2023pyp}.

The former is described by the following 4-dimensional Lagrangian, written in Minkowski space:
\begin{align}\label{eq:uvlagthesm}
 \mathcal{L}_\text{4}^\text{SM} &= 
 -\frac{1}{4}W^I_{\mu\nu}W^{I\,\mu\nu}-\frac{1}{4}B_{\mu\nu}B^{\mu\nu}\\\nonumber
 &+\overline{q}\ii\slashed{D}q+\overline{l}\ii\slashed{D} l+\overline{u}\ii\slashed{D}u+\overline{d}\ii\slashed{D}d+\overline{e}\ii\slashed{D}e\\\nonumber
 &+ (D_\mu\phi)^\dagger (D^\mu\phi)+\mu^2|\phi|^2 -\lambda|\phi|^4- (\overline{q} \tilde{\phi} Y_u u + \overline{q}\phi Y_d d+\overline{l}\phi Y_e e + \text{h.c.})\,,
\end{align}
where we use $B$ and $W$ for the EW $U(1)_Y$ and $SU(2)_L$ gauge bosons, with gauge couplings  $g_1$ and $g_2$, respectively. The Higgs doublet is represented by $\phi$, with $\tilde{\phi}\equiv \ii\sigma_2\phi^*$, while $\sigma_2$ is the second Pauli matrix. We adopt the minus-sign convention for the covariant derivative.

For the SMEFT, we have:
\begin{align}\label{eq:uvlagsmeft}
    \mathcal{L}_4^\text{SMEFT} &= \mathcal{L}_4^\text{SM} +  \frac{1}{\Lambda^2}\bigg\lbrace c_{\phi} (\phi^\dagger\phi)^3 + c_{\phi \square} (\phi^\dagger\phi)\square(\phi^\dagger\phi) + c_{\phi D} (\phi^\dagger D^\mu\phi)^* (\phi^\dagger D_\mu\phi)\nonumber\\
    &+c_{\phi \psi_L}^{(1)} (\phi^\dagger i \overleftrightarrow{D}_\mu\phi)(\overline{\psi_L}\gamma^\mu \psi_L) + c_{\phi \psi_L}^{(3)} (\phi^\dagger i \overleftrightarrow{D}^I_\mu\phi)(\overline{\psi_L}\gamma^\mu\sigma^I \psi_L) + c_{\phi \psi_R} (\phi^\dagger i \overleftrightarrow{D}_\mu\phi)(\overline{\psi_R}\gamma^\mu \psi_R)\nonumber\\
    & + \left[c_{\phi ud} (\widetilde{\phi}iD_\mu\phi)(\overline{u_R}\gamma^\mu d_R) + c_{\psi_R\phi} (\phi^\dagger\phi)\overline{\psi_L}\widetilde{\phi}\psi_R +\text{h.c.} \right]\,,
\end{align}
where $\psi_R = u, d, e$ and $\psi_L = q, l$. 
That is, we only consider those effective interactions that arise at tree-level in UV completions of the SM. (4-fermions are also disregarded, because they do not appear in one-loop calculations of 3-dimensional parameters.)
For the remainder of the paper we will absorb the cut-off scale, $\Lambda$, into the WCs, turning them dimensionful.

In the 3-dimensional EFT, gauge bosons $X_{\mu\nu}$ split into temporal components $X_0$ and spatial ones $X_{rs}$. The most generic Lagrangian can be written as
\begin{align}
 \mathcal{L}_{\text{3}} = \mathcal{L}_{\text{3}}^{(2)} + 
 \mathcal{L}_{\text{3}}^{(3)} + 
 \mathcal{L}_{\text{3}}^{(4)} +
 \mathcal{L}_{\text{3}}^{(5)} +
 \mathcal{L}_{\text{3}}^{(6)} + \cdots
\end{align}
where $\mathcal{L}_{\text{3}}^{(n)}$ contains the operators that can be generated at order $\mathcal{O}(g^n)$, assuming the usual power counting $\mu^2\sim P^2\sim Y^2\sim\lambda\sim g^2$, where $P$ represents a physical momentum, as well as $c_{\phi\Box}\sim c_{\phi D}\sim c_{\phi\psi}\sim g^2$, $c_{\psi\phi}\sim g^3$ and $c_{\phi}\sim g^4$. Thus, the number $n$ coincides with the dimension of the corresponding operators in 4-dimensional units. In what follows, we disregard odd powers of $n$ because they are not generated in our computation, in agreement with previous findings in the literature~\cite{Moore:1995jv,Kajantie:1997ky}.

With a little abuse of notation, we use the same name for 4-dimensional fields and for their zero modes. With the help of \texttt{Basisgen}~\cite{Criado:2019ugp}, \texttt{Sym2Int}~\cite{Fonseca:2017lem} and \texttt{ABC4EFT}~\cite{Li:2022tec}, we obtain, in Euclidean form: 
\begin{equation}\label{eq:eftlag2}
 \mathcal{L}_{\text{3}}^{(2)} = m_\phi^2|\phi|^2 + \frac{1}{2}m_{B_0}^2 B_0^2 + \frac{1}{2}m_{W_0}^2 W_{0I} W_0^I\,;
\end{equation}
\begin{align}\label{eq:eftlag4}
    \mathcal{L}_{\text{3}}^{(4)} &= k_\phi(D_r\phi)^\dagger (D^r \phi) + \frac{k_{B_0}}{2} (D_r B_0) (D^r B_0) + \frac{k_{W_0}}{2} (D_r W_{0I})(D^r W_0^I)\nonumber\\
    &+\frac{k_B}{4} B_{rs} B^{rs} + \frac{k_W}{4}W_{rs}^I W^{rs}_I + \lambda_{\phi^4}|\phi|^4 + \lambda_{B_0^4} B_0^4 + \lambda_{\phi^2 B_0^2} |\phi|^2 B_0^2\nonumber\\
    &+ \lambda_{W_0^4} (W_0^IW_{0I})^2 + \lambda_{\phi^2 W_0^2} |\phi|^2 W_0^I W_{0I} + \lambda_{B_0^2 W_0^2}B_0^2 W_0^I W_0^I + \lambda_{\phi^2 B_0 W_0} B_0 \phi^\dagger \sigma^I\phi W_0^I\,;
\end{align}
%
%
and
\begin{equation}\label{eq:eftlag6}
    \mathcal{L}_\text{3}^{(6)} = \mathcal{L}_{\text{3},\phi}^{(6)}+\mathcal{L}_{\text{3},B}^{(6)} + \mathcal{L}_{\text{3},\phi B}^{(6)} + \mathcal{L}_{\text{3},W}^{(6)} + \mathcal{L}_{\text{3},\phi W}^{(6)} + \mathcal{L}_{\text{3},B W}^{(6)} + \mathcal{L}_{\text{3},\phi B W}^{(6)} \,,
\end{equation}
where
\begin{align}
    \mathcal{L}_{\text{3},\phi}^{(6)} &=\textcolor{gray}{r_{\phi^2 D^4} D^2\phi^\dagger D^2\phi} + c_{\phi^4D^2}^{(1)}|\phi|^2 \Box|\phi|^2 + c_{\phi^4 D^2}^{(2)}(\phi^\dagger D^r\phi)^\dagger (\phi^\dagger D_r\phi)+ \textcolor{gray}{r_{\phi^4 D^2}^{(3)} |\phi|^2 D_r\phi^\dagger D^r\phi}\nonumber\\
    & +  \textcolor{gray}{r_{\phi^4 D^2}^{(4)} |\phi|^2 D_r(\phi^\dagger\ii\overleftrightarrow{D}^r\phi)} + c_{\phi^6} |\phi|^6\,,
 \end{align}
 \begin{align}
    \mathcal{L}_{3,B}^{(6)} &= \textcolor{gray}{r_{B_0^2 D^4} (D^2 B_0)^2 } + 
    \textcolor{gray}{r_{B_0^4 D^2}B_0^2 (D_r B_0)(D^r B_0)} + c_{B_0^6} B_0^6  + \textcolor{gray}{r_{B^2 D^2} (D_r B^{rs}) (D^t B_{ts})} \nonumber\\
    & + c_{B_0^2 B^2}B_0^2 B_{rs}B^{rs}\,,
\end{align}
\begin{align}
    \mathcal{L}_{3,\phi B}^{(6)} &=  c_{\phi^2 B^2}|\phi|^2 B_{rs}B^{rs} + c^{(1)}_{\phi^2  B_0^2 D^2} D_r\phi^\dagger D^r\phi B_0^2 + \textcolor{gray}{r_{\phi^2 B_0^2 D^2}^{(2)} |\phi|^2 (D_r B_0) (D^r B_0)}  \nonumber\\
    &+ 
    \textcolor{gray}{r_{\phi^2 B_0^2 D^2}^{(3)} D_r|\phi|^2 B_0 D^rB_0} + c_{\phi^2 B_0^4} |\phi|^2 B_0^4 + c_{\phi^4 B_0^2} |\phi|^4 B_0^2 + \textcolor{gray}{r_{\phi^2 BD^2} D_r B^{rs} (\phi^\dagger\ii\overleftrightarrow{D}_s\phi)}\,,
\end{align}
\begin{align}
    \mathcal{L}_{3,W}^{(6)} &= c_{W^3}\, \epsilon_{I,J,K} W^{I r}_s W^{J t}_{r} W^{K s}_t + c_{W_0^6} (W_0^I W^I_{0})^3 + c^{(1)}_{W_0^2 W^2} (W_0^I W_0^I) W^{J rs} W^J_{rs} \nonumber \\
    &+c^{(2)}_{W_0^2 W^2} (W_0^I W_0^J) W^{I rs} W^J_{rs} + \textcolor{gray}{r_{W^2D^2} (D_r W^{I rs}) (D^t W^I_{ts})} + c^{(1)}_{W_0^4 D^2} W^I_0 W^I_0 (D_r W_0^J) (D^r W_0^J) \nonumber \\
    &+\textcolor{gray}{r^{(2)}_{W_0^4 D^2} W^I_0 W^J_0 (D_r W_0^I) (D^r W_0^J)} + \textcolor{gray}{r_{W_0^2 W D^2} \,\epsilon^{IJK} W^I_{rs} (D^r W_0^J)(D^s  W_0^K)} \nonumber \\
    &+\textcolor{gray}{r_{W_0^2D^4} (D^2 W_0^I)(D^2 W_0^I)}\,,
\end{align}
\begin{align}
    \mathcal{L}_{3,\phi W}^{(6)} &=   c_{\phi^2 W^2}|\phi|^2 W^I_{rs}W^{I rs} + c_{\phi^2 W_0^4} |\phi|^2 (W_0^I W_0^I)^2  + c^{(1)}_{\phi^4 W_0^2} |\phi|^4 W_0^I W_0^I \nonumber \\
    &+ c^{(2)}_{\phi^4 W_0^2} (\phi^\dagger \sigma^I \phi) (\phi^\dagger \sigma^J \phi) W_0^I W_0^J + c^{(1)}_{\phi^2 W_0^2 D^2} |\phi|^2 (D_r W^I)(D^r W^I) \nonumber \\
    &+ c^{(2)}_{\phi^2 W_0^2 D^2}\, 
    \epsilon^{IJK} (\phi^\dagger \sigma^I \mathrm{i} \overleftrightarrow{D}_r \phi) W_0^J D^r W^K_0+ \textcolor{gray}{ r^{(3)}_{\phi^2 W^2_0 D^2} \phi^\dagger ( -\mathrm{i} \overleftrightarrow{D}_r) \phi (D^r W_0^I) W_0^I} \nonumber \\
    &+\textcolor{gray}{ r^{(4)}_{\phi^2 W^2_0 D^2}  \left[ D^2(\phi^\dagger)\phi + \phi^\dagger D^2\phi \right] W_0^I W_0^I } + \textcolor{gray}{ r^{(5)}_{\phi^2 W^2_0 D^2} D^2 (W_0^I) W_0^I |\phi|^2} \nonumber \\ 
    &+ \textcolor{gray}{r^{(6)}_{\phi^2 W^2_0 D^2} \, \epsilon^{IJK}D^2 (W_0^I) W_0^J \phi^\dagger \sigma^K \phi} + \textcolor{gray}{r_{\phi^2 W D^2}  D_s(W^{I rs}) \phi^\dagger \sigma^I \mathrm{i} \overleftrightarrow{D}_r\phi }\,,
\end{align}
\begin{align}
    \mathcal{L}_{3,BW}^{(6)} &= c_{B_0^2 W^2 } B_0^2 W^{I rs} W^I_{rs} + c_{B^2 W_0^2 } W_0^I W_0^I B^{rs} B_{rs} + c_{B_0^4 W_0^2} B_0^4 W_0^I W_0^I+ c_{B_0^2 W_0^4} B_0^2 (W_0^I W_0^I)^2 \nonumber \\ 
    &+ c^{(1)}_{B_0^2 W_0^2 D^2} B_0^2 D^r (W_0^I) D_r (W_0^I) + c_{B_0 B W_0 W} B_0 B^{rs} W_0^I W^I_{rs} \nonumber \\
    &+c_{B_0 W_0 W D^2} D^r(B_0) D^s(W_0^I) W^I_{rs}  + \textcolor{gray}{r^{(2)}_{B_0^2 W_0^2 D^2} W_0^I W_0^I D_r(B_0)D^r(B_0)}\nonumber \\
    &+ \textcolor{gray}{r^{(3)}_{B_0^2 W_0^2 D^2} B_0 D_r (B_0) D^r(W_0^I) W_0^I }\,,
\end{align}
\begin{align}
    \mathcal{L}_{3,\phi BW}^{(6)} &= c_{\phi^2 B_0^2 W_0^2} |\phi|^2 W_0^I W_0^I B_0^2 + c^{(1)}_{\phi^2 B_0 W_0 D^2} B_0 W_0^I D^r(\phi^\dagger) \sigma^I D_r(\phi) \nonumber \\
    &+ c^{(2)}_{\phi^2 B_0 W_0 D^2} B_0 D^r(W_0^I) D_r (\phi^\dagger \sigma^I \phi) +c_{\phi^2 B W} \phi^\dagger \sigma^I \phi W^I_{rs} B^{rs} \nonumber \\
    &+ c_{\phi^2 B_0^3 W_0} \phi^\dagger \sigma^I \phi W_0^I B_0^3 +  c_{\phi^2 B_0 W_0^3} \phi^\dagger \sigma^I \phi W_0^I  W_0^J W_0^J B_0 + c_{\phi^4 B_0 W_0}  \phi^\dagger \sigma^I \phi  |\phi|^2 W_0^I B_0 \nonumber \\
    &+\textcolor{gray}{r^{(3)}_{\phi^2 B_0 W_0 D^2} B_0 D^r(W_0^I) (\phi^\dagger \sigma^I\mathrm{i} \overleftrightarrow{D}_r\phi) } + \textcolor{gray}{r^{(4)}_{\phi^2 B_0 W_0 D^2}B_0 D^2(W_0^I) \phi^\dagger \sigma^I \phi } \nonumber \\
    &+\textcolor{gray}{r^{(5)}_{\phi^2 B_0 W_0 D^2} B_0 W^I_0 \left[\phi^\dagger \sigma^I D^2\phi + D^2(\phi^\dagger) \phi\right]} +\textcolor{gray}{r^{(6)}_{\phi^2 B_0 W_0 D^2} B_0 (W_0^I) (\phi^\dagger \sigma^I\mathrm{i} \overleftrightarrow{D}^2\phi) }\,.
\end{align}

Finally, we include the following operators with no 4-dimensional analog, namely involving the 3-dimensional Levi-Civita tensor (others are absent in our calculation):
%
%
%
%
%
%
\begin{align}\label{eq:rareoperators}
    \mathcal{L}^{DX}&= c_{\phi^2 W_0 W D}(\phi^\dagger \ii\overleftrightarrow{D}_r \phi) W_0^I \widetilde{W}_I^r 
+ c_{\phi^2 W_0 B D}(\phi^\dagger \sigma^I  \ii\overleftrightarrow{D}_r \phi) W_0^I \widetilde{B}^r \nonumber\\
&+ c_{\phi^2 B_0 W D}(\phi^\dagger \sigma^I  \ii\overleftrightarrow{D}_r \phi) B_0 \widetilde{W}_I^r 
%
+ c_{\phi^2 B_0 B D}(\phi^\dagger \ii\overleftrightarrow{D}_r \phi)B_0 \widetilde{B}^r \,.
%
\end{align}
where $\tilde{X}^r=\epsilon^{rst}X_{st}$. Some of these operators were studied in Ref.~\cite{Kajantie:1997ky} in the context of parity violation. Therein, other operators are also considered (such as those including only gauge bosons) but we neglect them as they are not generated in our work.

Operators in \textcolor{gray}{gray}, whose WCs we name with $r$, are redundant; they can be removed in favor of shifts on the WCs named with $c$ using field redefinitions. For brevity, we highlight here those shifts that will be relevant for later discussion:
\begin{align}\label{eq:redefinitions}
    m_\phi^2 &\to m_\phi^2 + m_\phi^4 r_{\phi^2 D^4}\,,\\
    \lambda_\phi &\to \lambda_\phi-g_2^2 m_\phi^2 r_{W^2D^2}+4\lambda_\phi m_\phi^2 r_{\phi^2 D^4} + 2g_2 m_\phi^2 r_{\phi^2 W D^2} + m_\phi^2 r_{\phi^4 D^2}^{(3)}\,,\\
    c_{\phi^6} &\to c_{\phi^6}-2g_2^2\lambda_\phi r_{W^2 D^2}+4 g_2\lambda_\phi r_{\phi^2 W D^2}+4\lambda_\phi^2 r_{\phi^2 D^4}-2\lambda_\phi r_{\phi^4 D^2}^{(3)}\,.
\end{align}

We refer to the set of operators in which redundant ones are removed following these shifts as \textit{physical basis}.

\section{Results}
\label{sec:results}
In order to determine the static 3-dimensional limit of the Lagrangian in Eqs.~\eqref{eq:uvlagthesm} and~\eqref{eq:uvlagsmeft}, we first compute the hard region of 1-particle irreducible off-shell correlators, namely Green's functions in the regime $Q^2\sim(\pi T)^2\gg P^2\sim\mu^2$, where $Q$ and $P$ refer to loop and external momenta, respectively. We do so by iterating the following identity:
\begin{equation}
    \frac{1}{(Q+P)^2+\mu^2} = \frac{1}{(Q^2+\mu^2)}\left[1-\frac{P^2+2 Q\cdot P+\mu^2)}{(Q+P)^2+\mu^2}\right]\,.
\end{equation}

We subsequently split 4-momenta into temporal and spatial components, $P^\mu = (P^0,\vec{P})$. For external momenta, $P^0$ vanishes. The result can be projected onto the EFT defined by Eqs.~\eqref{eq:eftlag2}--\eqref{eq:eftlag6}. We work in dimensional regularization with space-time dimension $D=1+d$, with $d=3-2\epsilon$, and employ the background-field method~\cite{Abbott:1981ke} in the Feynman gauge~\footnote{This implies that gauge couplings do not have to be matched separately. Likewise, wave-function renormalization factors are different from those derived without the background-field method; but matching conditions agree upon normalizing canonically all fields.}; see section~\ref{sec:crosschecks} though for comments in general $R_\xi$ gauge. We use \texttt{Feynrules}~\cite{Alloul:2013bka} for the generation of Feynman rules (with the modifications explained in Ref.~\cite{Chala:2024xll} to account for Euclidean space-time), as well as \texttt{Feynarts}~\cite{Hahn:2000kx} and \texttt{FeynCalc}~\cite{Shtabovenko:2016sxi} for the computation of Feynman diagrams and amplitude manipulations.

We write our results in terms of the sum-integrals $I_{\alpha}\equiv I_{\alpha,0,0}$, that we obtain from the following expressions for bosons and fermions respectively:
\begin{align}
    I_{\alpha,\beta,\delta}^b &= 
    \sumint_{Q} \frac{Q_0^{2\beta}|\vec{Q}|^{2\delta}}{Q^{2\alpha}}\\
    &= \frac{(e^{\gamma_E}\bar{\mu}^2)^\epsilon}{8\pi^2}\frac{\Gamma(\alpha-\frac{d}{2}-\delta) \Gamma(\frac{d}{2}+\delta) \zeta (2\alpha-2\beta-2\delta-d)}{\Gamma(\frac{1}{2})\Gamma(\alpha)\Gamma(\frac{d}{2})} (2\pi T)^{1+d-2\alpha+2\beta+2\delta}\,.\\
    I_{\alpha,\beta,\delta}^f &= \sumint_{\lbrace Q\rbrace} \frac{Q_0^{2\beta}|\vec{Q}|^{2\delta}}{Q^{2\alpha}}(2^{2\alpha-2 \beta-2\delta-d}-1) I_{\alpha,\beta,\delta}^b\,,
\end{align}
upon using the recursion relations~\cite{Brauner:2016fla}
\begin{align}
    I_{\alpha,\beta-1,\delta+1} &= -\frac{d/2 + \delta}{1 + 
    d/2 - \alpha + \delta} I_{\alpha,\beta,\delta}\,,\\
    I_{\alpha+1,\beta+1,0} &= \left(1-\frac{2}{2\alpha}\right) I_{\alpha,\beta,0}\,.
\end{align}
In the equations above, $\gamma_E$ is the Euler-Mascheroni constant and $\bar{\mu}$ represents the $\overline{\text{MS}}$ renormalization scale.

We provide the full set of relations determining how the 3-dimensional EFT WCs depend on the UV parameters in the ancillary files \texttt{matching\_fer.txt} and \texttt{matching\_bos.txt}, which include fermionic and bosonic modes contributions, respectively.
For illustration, we show below the results for the Higgs operators:
\begin{align}
    m_\phi^2 &= -\mu^2 + \frac{1}{4} \left\{ \left(-4 c_{\phi D}+8 c_{\phi\Box}\right) \left(I^b_2 \mu^4-I^b_1 \mu^2\right)+ \left[d \left(g_1^2+3
   g_2^2\right)+24 \lambda \right] I_1^b \right.\nonumber\\
   &\left.-\left(I_3^b \mu^4 - I_2^b\mu^2 \right)
   \left[ g_1^2+3 \left(g_2^2- 8 \lambda \right)\right]\right\}\nonumber\\
   &-2 \mathrm{Tr}\left[Y_e Y_e^\dagger + N_C Y_d Y_d^\dagger + N_C Y_u Y_u^\dagger\right] I_1^f\,, \\
    %
    %
    k_\phi &= 1+\left( c_{\phi D}-2 c_{\phi\Box}\right) \left(-I^b_2 \mu^2+I^b_1\right)-\frac16(g_1^2+3g_2^2)(3 I_2^b-2\mu^2I_3^b)\nonumber\\
    &+\mathrm{Tr}\left[Y_e Y_e^\dagger + N_C Y_d Y_d^\dagger + N_C Y_u Y_u^\dagger\right] I_2^f\,,\\
    %
    %
    %
    \frac{\lambda_{\phi^4}}{T} &= \lambda +\frac{1}{16}\left\{ -d(g_1^4+2g_1^2g_2^2+3g_2^4)I_2^b+8\lambda[(I_2^b-2\mu^2I_3^b)(g_1^2+3g_2^2-24\lambda)]\right\}\nonumber\\
    &-12 c_\phi(I_1^b-\mu^2I_2^b)+\frac14 c_{\phi D}\left[d(g_1^2+g_2^2)I_1^b-16\lambda(I_1^b-2\mu^2I_2^b)\right]\nonumber\\
    &+12c_{\phi \Box}\lambda(I_1^b-2\mu^2I_2^b) + \mathrm{Tr}\left[Y_e Y_e^\dagger Y_e Y_e^\dagger + N_C Y_d Y_d^\dagger Y_d Y_d^\dagger + N_C Y_u Y_u^\dagger Y_u Y_u^\dagger \right] I_2^f\nonumber \\ 
    &+2\mathrm{Tr}\left[N_C(c_{d\phi}Y_d^\dagger + c_{d\phi}^\dagger Y_d+c_{u\phi}Y_u^\dagger + Y_u c_{u\phi}^\dagger)\right]I_1^f\,,\\
    %
    %
    %
    r_{\phi^2 D^4} &= \frac{1}{6}(g_1^2+3g_2^2)I_3^b-\frac13 \mathrm{Tr}\left[Y_e Y_e^\dagger + N_C Y_d Y_d^\dagger + N_C Y_u Y_u^\dagger\right] I_3^f\,,
    \end{align}
\begin{align}
    \frac{c_{\phi^4 D^2}^{(1)}}{T} &= c_{\phi\Box}-\frac{1}{96}\left\{ (5+2d)g_1^4+(37+6 d)g_2^4+2g_1^2\left[(5+2d)g_2^2-64\lambda\right]-224g_2^2\lambda+320\lambda^2\right\}I_3^b\nonumber\\
    &+c_{\phi D}\left(\lambda-\frac34 g_2^2\right)I_2^b+\frac14 c_{\phi\Box}\left(-g_1^2+3g_2^2-56\lambda\right)I_2^b\nonumber\\
    &+\frac13 \mathrm{Tr}\left[Y_e Y_e^\dagger Y_e Y_e^\dagger + N_C Y_d Y_d^\dagger Y_d Y_d^\dagger + N_C Y_u Y_u^\dagger Y_u Y_u^\dagger\right] I_3^f+\frac12 N_C \mathrm{Tr}\left[Y_d^\dagger c_{d\phi}+Y_d a^\dagger_{d\phi}\right.\nonumber\\&\left.+2 Y_d Y_d^\dagger(-c_{\phi d}
    +a^{(1)}_{\phi q}+a^{(3)}_{\phi q}) + Y_u^\dagger c_{u\phi}+Y_u a^\dagger_{u\phi} + 2 Y_u Y_u^\dagger(-a^{(1)}_{\phi q}+a^{(3)}_{\phi q}+c_{\phi u})\right]I_2^f\nonumber\\
    &+\frac12\mathrm{Tr}\left[Y_e^\dagger c_{e \phi}+Y_e a^\dagger_{e\phi}+2 Y_e Y_e^\dagger(-c_{\phi e}+a^{(1)}_{\phi\ell}+a^{(3)}_{\phi\ell})\right]I_2^f\,,
\end{align}
    \begin{align}
    \frac{c_{\phi^4 D^2}^{(2)}}{T} &= c_{\phi D}+\frac{1}{12}g_1^2\left[(15-2d)g_2^2+40\lambda\right]I_3^b-\frac14c_{\phi D} (g_1^2+21g_2^2+24\lambda)I_2^b-3c_{\phi\Box}g_1^2 I_2^b\nonumber\\
    &-\frac23 \mathrm{Tr}\left[Y_e Y_e^\dagger Y_e Y_e^\dagger + N_C Y_d Y_d^\dagger Y_d Y_d^\dagger + N_C Y_u Y_u^\dagger Y_u Y_u^\dagger-2N_CY_d Y_u Y_d^\dagger Y_u^\dagger\right] I_3^f \nonumber\\
    &- 2 N_C \mathrm{Tr}\left[ 2 Y_u^\dagger Y_u\left( c_{\phi q}^{(1)}-c_{\phi u}\right) - Y_u^\dagger Y_d c_{\phi u d}^\dagger -  c_{\phi u d} Y_d^\dagger Y_u + 2 Y_d^\dagger Y_d (c_{\phi d}-a^{(1)}_{\phi q})\right]I_2^f \nonumber\\
    &-4\mathrm{Tr}\left[Y_e^\dagger Y_e (c_{\phi e}-c_{\phi\ell}^{(1)})\right]\,,
    %
    %
    %
    %
    %
    %
    %
    %
    %
    %
    %
    %
\end{align}
\begin{align}
    \frac{r_{\phi^4 D^2}^{(3)}}{T} &= \frac{1}{12}\left\{5g_1^4+g_1^2[(-5+2d)g_2^2+8\lambda]+4(g_2^4+16g_2^2\lambda+8\lambda^2)\right\}I_3^b\nonumber\\
    &-\frac12 c_{\phi D} (3g_1^2-3g_2^2+4\lambda) I_2^b + c_{\phi\Box}(4\lambda-3g_2^2)I_2^b\nonumber\\
    &-\frac43\mathrm{Tr}\left[Y_e Y_e^\dagger Y_e Y_e^\dagger + N_C Y_d Y_d^\dagger Y_d Y_d^\dagger + N_C Y_u Y_u^\dagger Y_u Y_u^\dagger-2N_CY_d Y_u Y_d^\dagger Y_u^\dagger\right] I_3^f\nonumber\\
    &-N_C\mathrm{Tr}\left[Y_d^\dagger c_{d\phi}+Y_d a^\dagger_{d\phi}-4Y_d Y_d^\dagger a^{(3)}_{\phi q}+ 2 Y_u Y_d^\dagger c_{\phi u d}+ 2  c_{\phi u d}^\dagger Y_dY_u^\dagger +Y_u^\dagger c_{u\phi}+Y_u a^\dagger_{u\phi}\right.\nonumber\\
    &\left.-4 Y_u Y_u^\dagger a^{(3)}_{\phi q} \right]I_2^f-\mathrm{Tr}[Y_e^\dagger c_{e\phi}+c_{e\phi}^\dagger Y_e - 4 Y_e Y_e^\dagger a^{(3)}_{\phi \ell}]I_2^f\,,\\
    \frac{r_{\phi^4 D^2}^{(4)}}{T} &= -\mathrm{Im}\mathrm{Tr}\left[c_{e\phi}Y_e^\dagger + N_C\left(c_{d\phi}Y_d^\dagger-c_{u\phi}Y_u^\dagger\right)\right]I_2^f\,,\\
    \frac{c_{\phi^6}}{T^2} &= -c_{\phi}+\frac{1}{48}\left[d(g_1^6+3g_1^4g_2^2+3g_1^2g_2^4+3g_2^6)-48\lambda^2(g_1^2+3g_2^2-40\lambda)\right]I_3^b \nonumber\\
    &-\frac34 c_{\phi}\left(g_1^2+3g_2^2-72\lambda\right)I_2^b+\frac{1}{48}c_{\phi D}\left[960\lambda^2-6 d (g_1^2+g_2^2)^2\right]I_2^b-72c_{\phi\Box}\lambda^2 I_2^b\nonumber\\
    &-\frac23\mathrm{Tr}\left[Y_e Y_e^\dagger Y_e Y_e^\dagger Y_e Y_e^\dagger + N_C Y_d Y_d^\dagger Y_d Y_d^\dagger Y_d Y_d^\dagger + N_C Y_u Y_u^\dagger Y_u Y_u^\dagger Y_u Y_u^\dagger\right]I_3^f\nonumber\\
    &-2N_C\mathrm{Tr}\left[Y_d Y_d^\dagger (Y_d c_{d\phi}^\dagger+c_{d\phi}Y_d ^\dagger) + Y_u Y_u^\dagger (Y_u c_{u\phi}^\dagger+c_{u\phi}Y_u ^\dagger)\right]I_2^f\nonumber\\
    &-2\mathrm{Tr}\left[Y_e Y_e^\dagger (Y_e c_{e\phi}^\dagger+c_{e\phi}Y_e ^\dagger)\right]I_2^f\,,
\end{align}
where $N_C$ stands for the number of fermion colors. Note that the WC $r_{\phi^4 D^2}^{(4)}$ corresponds to a CP-violating operator and as such is not generated by considering only SM interactions (the SM CP-violating phase does not appear at this order) but it is instead induced by new sources of CP-violation captured by the SMEFT; the result is proportional to the leading invariants introduced in Ref. \cite{Bonnefoy:2021tbt}.

\section{Cross checks}
We have compared our results to order $\mathcal{O}(g^4)$ with previous works in the literature~\cite{Brauner:2016fla,Croon:2020cgk}, with full agreement. Moreover, we have verified that the UV divergences arising in the computation of the hard region of the one-loop processes involving SMEFT operators match those obtained at zero temperature~\cite{Chala:2021pll}. Likewise, we have checked that 3-dimensional Ward-identities hold, and that certain Green's functions for which there is no 3-dimensional operator vanish; e.g. the one with six $B_r$ vectors, despite the fact that the counterpart with temporal components is non-zero.

Beyond this, there is little to cross-check against existing results, as we discuss further below.

\label{sec:crosschecks}

\subsection{Comparison with Ref.~\cite{Moore:1995jv}}
To our knowledge, it is only in Ref.~\cite{Moore:1995jv} that a 
non-negligible part of our computations were performed before. However, only loops of fermions and only within the SM were included, and $g_1$ (and correspondingly $B$ and $B_0$ operators) was neglected. The results of Ref.~\cite{Moore:1995jv} are expressed in terms of operators different from ours. Explicitly, using $a$ for the WCs therein:
\begin{align}
    \mathcal{L}_3^{(6)} &= a_{\phi^6} |\phi|^6 + a_{\phi^2 W_0^4} |\phi|^2 (W_0^I W_0^I)\nonumber\\
    &+ a_{\phi^4 D^2}^{(1)} |\phi|^2 (D_r\phi)^\dagger)(D^r\phi) + a_{\phi^4 D^2}^{(2)} (\phi^\dagger D_r\phi) (D^r\phi)^\dagger \phi + a_{\phi^4 D^2}^{(3)} \partial_r(\phi^\dagger\phi) \partial^r (\phi^\dagger\phi)\nonumber\\
    &+ a_{\phi^2 D^4} D^2(\phi)^\dagger D^2\phi+ a_{\phi^2 W_0^2 D^2}^{(1)} W_0^I W_0^I (D_r\phi)^\dagger D^r\phi + a_{\phi^2 W_0^2 D^2}^{(2)} (D_r W_0^I) (D^r W_0^I) |\phi|^2\nonumber\\
    &+ a_{W_0^4 D^2} D_r(W_0^I W_0^I)D^r (W_0^I W_0^I) + a_{W_0^2 D^4} (D^2 W_0^I) (D^2 W_0^I)+ a_{\phi^2 W^2} W_{rs}^I W^{I\,rs} |\phi|^2\nonumber\\
    & + a_{W_0^2 W^2}^{(1)} W_0^I W_0^I W_{rs}^J W^{J\,rs} + a_{W_0^2 W^2}^{(2)} W_0^I  W_{rs}^I W_0^J W^{J\,rs}\nonumber\\
    &+ a_{W^3} \epsilon^{IJK} W_{rs}^I W_{st}^J W^{K\,tr} + a_{W^2 D^2} (D^r W_{rs}^I) (D_t W^{I\,ts})\,.
\end{align}

This Lagrangian can be projected onto our Green's basis simply using integration by parts and certain algebraic identities. In practice, we do so by equating off-shell amplitudes at tree level~\cite{Chala:2021cgt}. We obtain the following relations:
\begin{align}
\label{eq:moorei}
c_{\phi^4 D^2}^{(1)} &= -a_{\phi^4 D^2}^{(3)}\,,\\
c_{\phi^4 D^2}^{(2)} &= a_{\phi^4 D^2}^{(2)}\,,\\
r_{\phi^4 D^2}^{(3)} &= a_{\phi^4 D^2}^{(1)}\,,\\
c_{\phi^6} &= a_{\phi^6}\,,\\
c_{W^3} &= a_{W^3}\,,\\
r_{W^2 D^2} &= a_{W^2 D^2}\,,\\
r_{W_0^2 D^4} &= a_{W_0^2 D^4}\,,\\
r_{W_0^4 D^2}^{(2)} &= 4 a_{W_0^4 D^2}\,,\\
%
%
%
c_{\phi^2 W^2} &= a_{\phi^2 W^2}\,,\\
c_{\phi^2 W_0^4} &= a_{\phi^2 W_0^4}\,,\\
c_{\phi^2 W_0^2 D^2}^{(1)} &= a_{\phi^2 W_0^2 D^2}^{(1)}+a_{\phi^2 W_0^2 D^2}^{(2)}\,,\\
r_{\phi^2 W_0^2 D^2}^{(4)} &= -\frac{1}{2}a_{\phi^2 W_0^2 D^2}^{(1)}\,\\
r_{\phi^2 W_0^2 D^2}^{(5)} &= a_{\phi^2 W_0^2 D^2}^{(1)}\,\\
c_{W_0^2 W^2}^{(1)} &= a_{W_0^2 W^2}^{(1)}\,,\\
c_{W_0^2 W^2}^{(2)} &= a_{W_0^2 W^2}^{(2)}\,;
\label{eq:mooref}
\end{align}
all others vanish.

Using the matching conditions in Ref.~\cite{Moore:1995jv} and Eqs.~\eqref{eq:moorei}--\eqref{eq:mooref}, we obtain that all results agree with ours (in the limit of no bosons and $g_1\to 0$) with the exception of the following ones: $c_{\phi^2 W^2}$, $c_{W_0^2 W^2}^{(1,2)}$, $r_{\phi^2 W D^2}$ and $r_{W_0^2 WD^2}$.

It seems to us that this discrepancy lies in that 3-point functions with a single $W$  were not computed in Ref.~\cite{Moore:1995jv}. This amounts to assuming that $r_{\phi^2 WD^2}$ and $c_{W_0^2 W D^2}$ vanish. Since these operators enter into the matching conditions for Green's functions relevant for fixing $c_{\phi^2 W^2}$ (e.g. $\phi \phi W W$), as well as into those relevant for $c_{W_0^2 W^2}^{(1)}$ and $c_{W_0^2 W^2}^{(2)}$ (e.g. $W_0 W_0 W W$), with one of the $W$s coming from the covariant derivative, then the matching conditions of these WCs get spoiled.
The fact that operators like $r_{W_0^2 W D}$ are absent in Ref.~\cite{Moore:1995jv} seems to contradict also similar calculations in hot QCD~\cite{Chapman:1994vk,Laine:2018lgj}. Of course, nothing of the above changes any of the conclusions drawn in Ref.~\cite{Moore:1995jv}.

\subsection{Gauge independence}
The results presented in section~\ref{sec:results} were obtained in the Feynman gauge. However, as a consistency check of our computations, we have also performed some calculations in arbitrary $R_\xi$ gauge, and tested that the WCs in the physical basis (which themselves are directly related to physical quantities) are independent of $\xi$~\footnote{Right before the submission of this work, Ref.~\cite{Bernardo:2025vkz} appeared on the arXiv and shows gauge dependence canceling in the Abelian Higgs model. See also Ref.~\cite{Balui:2025kat} for a recent discussion of gauge dependence in the functional approach.}. 

We have paid particular attention to the computation of $c_{\phi^6}$, since Ref.~\cite{Croon:2020cgk} found  that the one-loop matching of this parameter was seemingly gauge dependent. As it was also correctly anticipated there, the problem stems from the fact that certain redundant interactions contribute to this parameter upon field redefinitions; see Eq.~\eqref{eq:redefinitions}. Indeed, in $R_\xi$ gauge and considering only SM contributions ensuing from bosonic loops, the matching conditions are given by:
\begin{align}
    c^G_{\phi^6} &= \frac{1}{48} \left[ d (g_1^6+3g_1^4g_2^2+3g_1^2g_2^4+3g_2^6) - 48\lambda^2(g_1^2\xi + 3 g_2^2 \xi)+1920\lambda^3\right]I^b_3\,, \nonumber \\
    r^{(3)}_{\phi^4D^2} &= \frac{1}{12} \left[(2d-5)g_1^2 g_2^2+4g_1^2\lambda(5-3\xi) + 5 g_1^4-g_2^4 (\xi^2+10\xi-15)+4g_2^2\lambda (25-9\xi)+32\lambda^2  \right]I^b_3\,,\nonumber
\end{align}
\begin{align}
    r_{\phi^2 D^4} &=-\frac{1}{12}(3g_1^2\xi-5g_1^2+9g_2^2\xi-15g_2^2)I^b_3 \,,\nonumber\\
    r_{\phi^2WD^2} &= -\frac{1}{12} (g_1^2 g_2 + g_2^3\xi^2+8g_2^3\xi-19g_2^3)I^b_3\,,\nonumber\\
    r_{W^2D^2} &=-\frac{1}{60} g_2^2\left[2d+5\xi(\xi+6)-116\right]I^b_3\,;
\end{align}
where the $G$ superscript in $c^G_{\phi^6}$ serves to specify that the WC is defined in the Green's basis and we set the gauge parameter associated to all gauge bosons to be the same, $\xi$, for simplicity. The gauge-dependent result for $c^G_{\phi^6}$ agrees with what was obtained in Ref.~\cite{Croon:2020cgk}.

Hence, following Eq.~\eqref{eq:redefinitions}, we obtain that
\begin{align}
    c_{\phi^6} &= \frac{1}{240} \left\{ -8\lambda\left[5(2d-3)g_1^2g_2^2+(1-2d)g_2^4+25g_1^4\right] + 5 d (g_1^6+3g_1^4g_2^2+3g_1^2g_2^4+3g_2^6)\right.\nonumber\\
    &\left.-400\lambda^2(g_1^2+7g_2^2)+8320\lambda^3\right\}I_3^b \,,
\end{align}
which, as expected given that $c_{\phi^6}$ is related to a physical quantity, is independent of $\xi$.
The fact that more coefficients are necessary to be considered (besides just computing the correlator of $\phi^6$) not only cancels the unphysical gauge-dependence but also changes the numerical factors of $c^G_{\phi^6}$ and introduces new terms, such as those scaling as $\lambda g_{1,2}^4$ which were absent for $c^G_{\phi^6}$. 

At $\mathcal{O}(g^6)$, we also obtain that $m_\phi^2$ and $\lambda_{\phi}$ are gauge-dependent in the Green's basis; after canonically normalizing the Higgs field, in $R_\xi$ gauge we obtain:
\begin{align}
    m^G_{\phi} &= \frac{1}{12}\left\{ 3 \left[d (g_1^2+3g_2^2)+24\lambda\right]I^b_1- 9 m^2_\phi (g_1^2+3g_2^2-8\lambda)I_2^b + \right. \nonumber\\
&\left.+m_\phi^4 \left[g_1^2(3\xi-10)+g_2^2(9\xi-30)+72\lambda\right]I_3^b\right\}\,.
\end{align}

This dependence is once again canceled when going to the physical basis following the shifts of Eqs.~\eqref{eq:redefinitions}. The same cancellation is obtained for $\lambda_{\phi}$. 

\section{Discussion and outlook}
\label{sec:outlook}
We have computed the one-loop effective action describing the high-temperature limit of the EW sector of the SM and of the SMEFT at the soft scale $\sim gT$ to order $\mathcal{O}(g^6)$. To this aim, we have worked out for the first time a basis of independent 3-dimensional Green's functions involving the zero-temperature modes of the Higgs, the EW gauge bosons and their temporal components, including the removal of some unphysical terms via field redefinitions.
Our work extends previous results in the literature, that include the full $\mathcal{O}(g^4)$ as well as the one-loop $\mathcal{O}(g^6)$ corrections in the SM only, and elucidates the seemingly gauge-dependence of certain physical parameters.

Our work is technical in nature, but it finds immediate applications to different aspects of thermal field theory. To highlight one, we very roughly estimate the size of the higher-order corrections in the determination of PT transition parameters in the SMEFT when the only non-vanishing dimension-6 WC is $c_\phi$. The 3-dimensional tree-level scalar potential, ignoring light Yukawas and for $\bar{\mu}=4 \pi T$, reads:
\begin{align}
    V_3 &= \left[-\mu^2+\frac{1}{16}(g_1^2+3 g_2^2+8\lambda_\phi+4 Y_t^2)T^2\right]|\phi|^2 \nonumber\\
    &+ \left[\lambda_\phi T-\frac{c_\phi}{\Lambda^2} T^3+\frac{T}{128\pi^2} (g_1^4+2 g_1^2 g_2^2+3 g_2^4)\right]|\phi|^4 -c_\phi T^2|\phi|^6 + \mathcal{O}(g^6)\,.
\end{align}
\begin{figure}[t]
    \includegraphics[width=0.32\columnwidth]{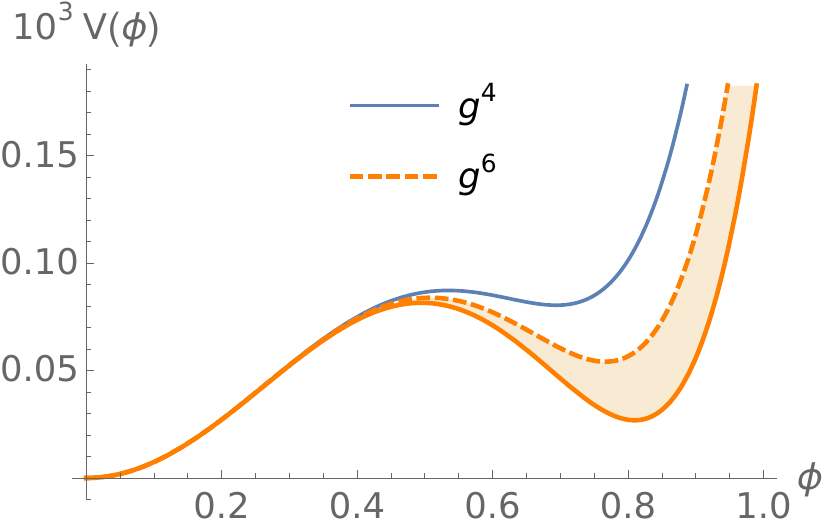}
    \includegraphics[width=0.32\columnwidth]{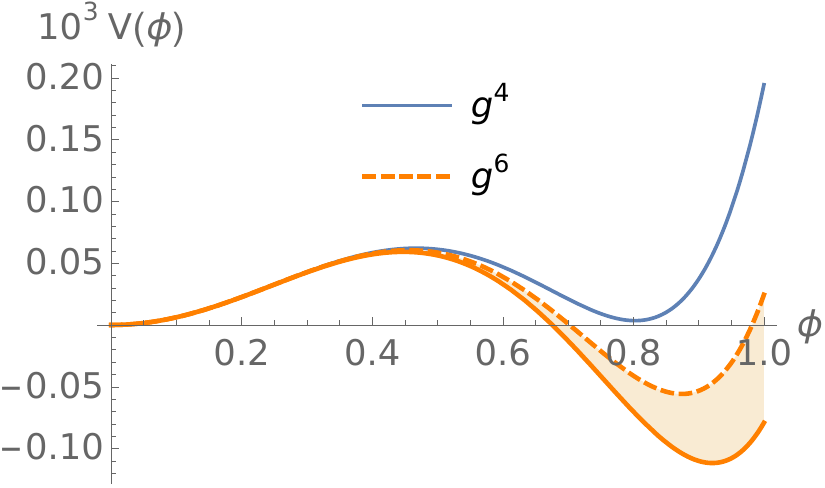}
    \includegraphics[width=0.32\columnwidth]{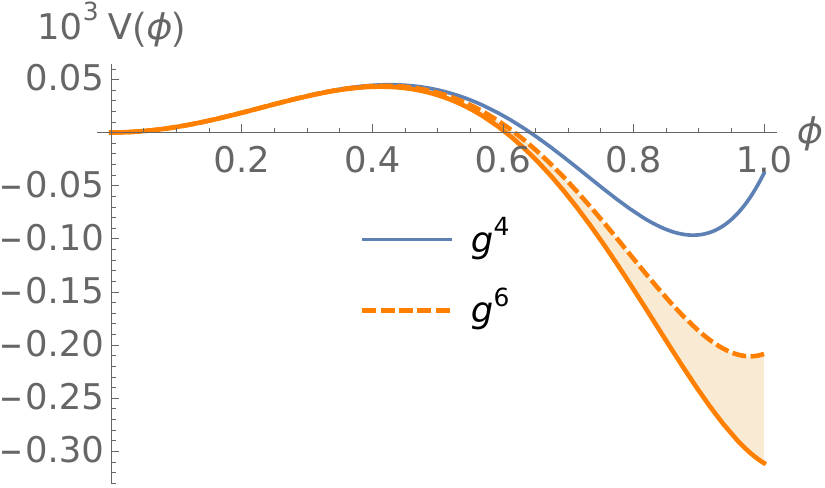}
    \caption{\it Scalar potential at different orders in $g$ at temperatures 10\% above (left), at (middle) and 10\% below (right) the critical temperature $T_c$ computed at $\mathcal{O}(g^4)$ for the case $c_\phi$ = 2.9; all in TeV units. The orange band encodes the difference between including (solid) and neglecting (dashed) redundant operator contributions to $c_{\phi^6}$. For the latter case, we consider the results in the Feynman gauge.}
    %
    %
    \label{fig:potential}
\end{figure}

\begin{figure}[t]
    \includegraphics[width=0.49\columnwidth]{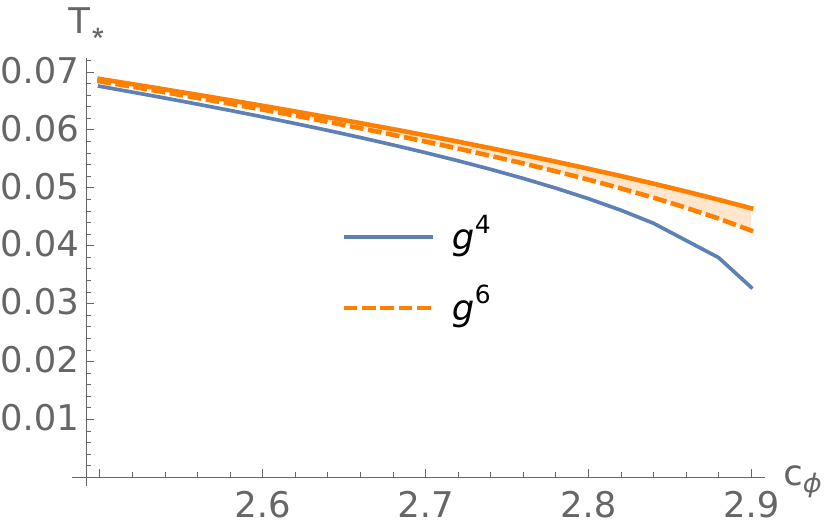}
    \includegraphics[width=0.49\columnwidth]{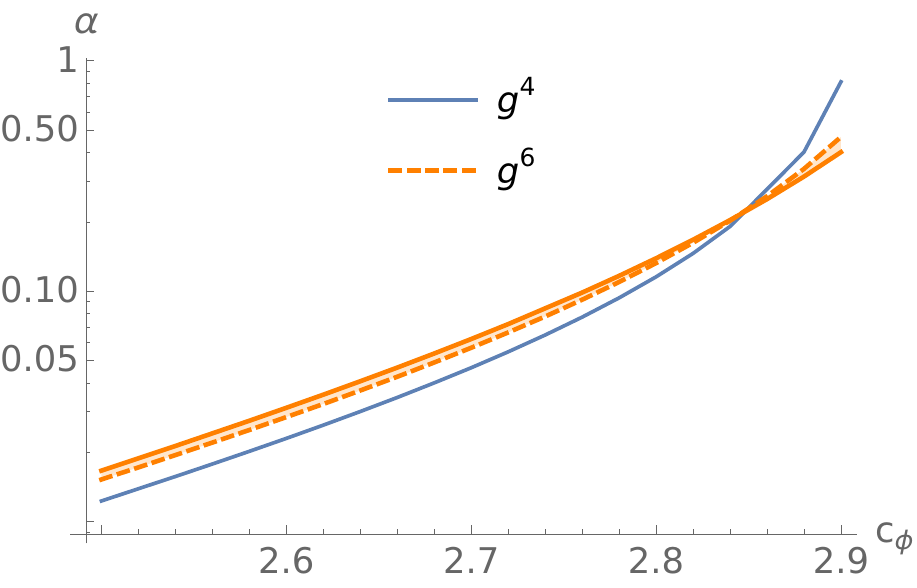}
    \includegraphics[width=0.49\columnwidth]{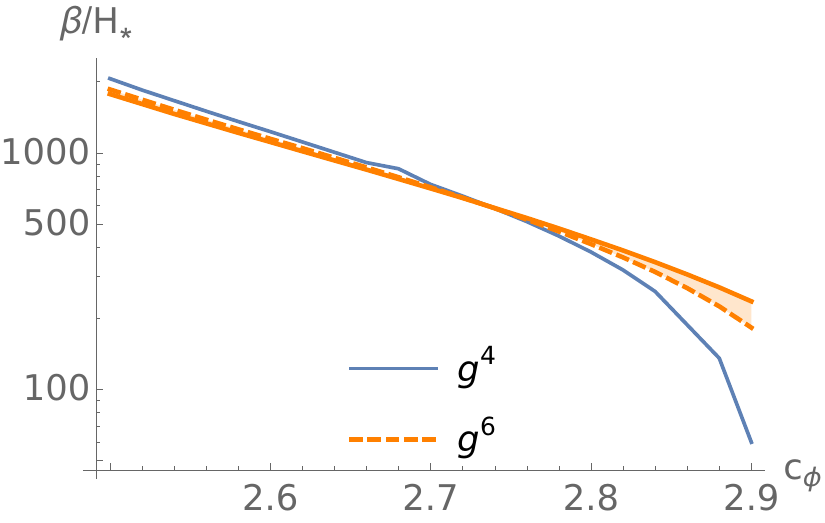}
    \includegraphics[width=0.49\columnwidth]{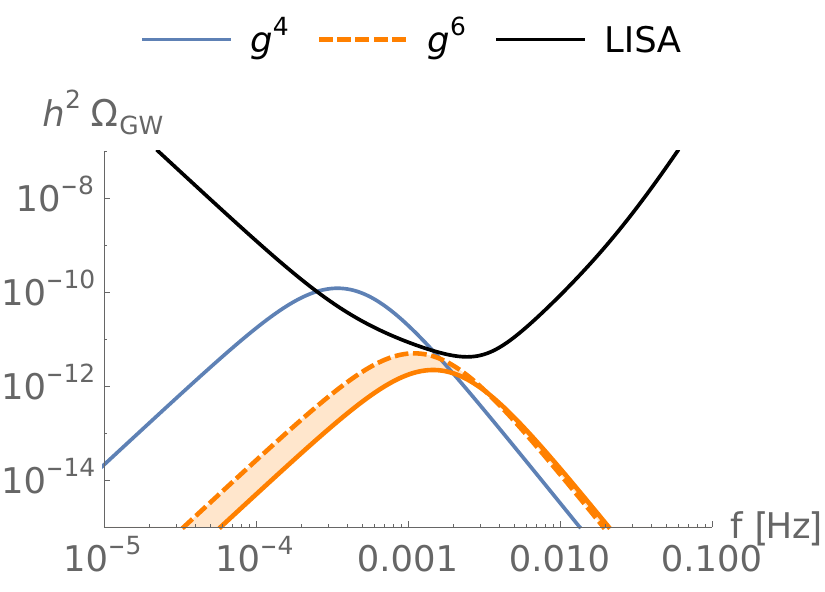}
    \caption{\it Nucleation temperature $T_*$ (top left), latent heat $\alpha$ (top right), inverse-duration time $\beta/H_*$ (bottom left) of the PT taking place in the SMEFT as a function of $c_\phi$ and GW spectrum at $c_\phi=2.9$ (bottom right), at different orders in $g$ for $\Lambda=1$; all in TeV units. The orange band encodes the difference between including (solid) and neglecting redundant (dashed) operator contributions to $c_\phi$.}
    \label{fig:PTparameters}
\end{figure}

In Fig.~\ref{fig:potential}, we depict this quantity at temperatures around the critical temperature $T_c$, defined as that at which the $\mathcal{O}(g^4)$ scalar potential presents two degenerated minima. In Fig.~\ref{fig:PTparameters}, we show the nucleation temperature, the strength and the inverse duration time of the PT computed with the help of \texttt{FindBounce}~\cite{Guada:2020xnz}, all derivative interactions being neglected~\footnote{A proper computation of the PT parameters should include the non-negligible effect of derivative interactions in the EFT, via appropriate perturbative functional analysis as worked out in Ref.~\cite{Chala:2024xll}; see also Refs.~\cite{Baacke:1995hw,Hirvonen:2021zej,Hua:2025fap}. It should also include higher-loop corrections~\cite{Gynther:2005dj,Ghisoiu:2012kn,Schroder:2012hm,Davydychev:2023jto}, to quartic and mass terms, as well as quantum corrections in the EFT~\cite{Baacke:1993ne,Dunne:2005rt,Ai:2018guc,Ai:2020sru,Ekstedt:2023sqc,Ai:2023yce,Matteini:2024xvg} (e.g. the effective potential). This is beyond of the scope of our work though, and our naive estimation suffices to demonstrate the importance of $\mathcal{O}(g^6)$ terms in the SMEFT PT.}. For each value of $c_\phi$, we fix $\mu^2$ and $\lambda$ to fit the Higgs vaccum expectation value $v\sim 246$ GeV and the mass $m_h\sim 125$ GeV at zero temperature.
We also show the GW spectrum for $c_{\phi}=2.9$, obtained with \texttt{PTPlot}~\cite{PTPlot,Caprini:2019egz} (for simplicity we have simply assumed the bubble wall velocity to be $v_w=1)$. The sensitivity curve of LISA is drawn too.
It is evident from the figure that $\mathcal{O}(g^6)$ corrections can modify the leading-order results by 50\,\%  in the strongest PTs. This reflects in sizable changes in the GW spectrum. We have also checked that non-vanishing values for $c_{\phi D}$ or $c_{u\phi}$ reduce the allowed range of values for $c_\phi$ where a PT takes place.
(We acknowledge that the large values of $c_\phi$ require a rather low cut-off. While the phenomenological viability of this has been discussed elsewhere~\cite{Chala:2018ari,Postma:2020toi,Hashino:2022ghd,Damgaard:2015con}, here we simply use it as a toy example.)

Our work paves the way for accurate computations of thermal parameters (and most precisely of PT and GW ones) in the SMEFT, and so within any extension of the SM in which there is a gap between the new physics and the EW scales, and provided that $T<\Lambda$. Still, achieving full $\mathcal{O}(g^6)$ requires a substantial amount of further computations. This includes, among others: (i) the extension of the 3-dimensional basis with gluonic operators~\cite{Chapman:1994vk,Laine:2018lgj} (they do not intervene in the leading soft-scale Higgs potential, but they contribute at one loop); (ii) the computation of relevant 2- and 3-loop matching corrections, which would then involve 4-fermion SMEFT (see Ref.~\cite{Gynther:2005dj} for results within the SM); (iii) 2- and 3-loop running within the 3-dimensional theory; (iv) the EFT at the ultrasoft scale, resulting from integrating out the temporal components of 3-dimensional gauge fields with Debye masses; (v) the computation of the 3-loop effective potential. We plan to address some of these points in future works.

\section*{Acknowledgments}
We are enormously grateful to Andreas Ekstedt for collaboration during the first stages of this project and for very detailed feedback on the manuscript. We thank A. Dashko, L. Gil, J. L\'opez-Miras, P. Olgoso and J. Santiago for useful discussions. We are particularly thankful to Z. Ren for helping us with the generation of certain Green's functions using an unpublished version of \texttt{ABC4EFT}~\cite{Li:2022tec}. We thank Javier Fuentes-Mart\'in, Javier L\'opez-Miras and Adri\'an Moreno-S\'anchez for cross-checking our results with a new version of \texttt{Matchete}~\cite{Fuentes-Martin:2022jrf}, and for spotting some wrong signs and missing terms in the first ancillary file. MC acknowledges support from the MCIN/AEI (10.13039/501100011033) and ERDF (grants
PID2021-128396NB-I00 and PID2022-139466NB-C21/C22), from the Junta de Andaluc\'ia grants FQM 101 and P21-00199 and from Consejer\'ia de Universidad, Investigaci\'on
e Innovaci\'on, Gobierno de Espa\~na and Uni\'on Europea -- NextGenerationEU under grants AST22 6.5 and CNS2022-136024, as well as from the RyC program under contract number RYC2019-027155-I. The work of GG is supported by the Deutsche Forschungsgemeinschaft under Germany’s Excellence Strategy EXC 2121 “Quantum Universe” -- 390833306, as well as by the grant 491245950.

\bibliographystyle{style} 

\bibliography{refs} 

\end{document}